\documentclass[aps,pre,reprint,twocolumn,superscriptaddress,showpacs,notitlepage]{revtex4-2}
\usepackage{amsmath}
\usepackage{amssymb}
\usepackage{emptypage}
\usepackage{graphicx}

\usepackage{xcolor}

\begin{document}

\title{Language dynamics within adaptive networks: \\
An agent-based approach of nodes and links coevolution}

\author{Christos Charalambous, David S\'anchez, Ra\'ul Toral}
\affiliation{Institute for Cross-Disciplinary Physics and Complex Systems IFISC (UIB-CSIC), Campus Universitat Illes Balears, E-07122 Palma de Mallorca, Spain.}
\begin{abstract}
Motivated by the dramatic disappearance of endangered languages observed in
recent years, a great deal of attention has been given to the
modeling of language competition in order to understand the factors
that promote the disappearance of a language and its unfolding dynamics.
With this in mind, we build on existing network models of
language competition in bilingual societies. These models deal with the interplay between the
usage of a language (link state) and the preference or attitude of the
speakers towards the language (node state). In this work, we allow for the case where agents have the freedom to adapt their
local interactions in accordance with their language preference. This
is modeled by introducing a local rewiring mechanism triggered by the dissatisfaction of an agent with its usage of a given language. Our numerical simulations show that permitting this freedom to agents likely results in linguistically segregated communities for small network sizes. However, for networks of sufficiently large size, the extinction of one of the languages is the most probable scenario. Furthermore, we analyze how the fraction of minority speakers changes with the system size and we find that this fraction grows as the total population increases, which is consistent with existing data.
Overall, the results of this work help us understand the impact of speakers' preferences and choices in the complex language landscape of bilingual societies. 
\end{abstract}

\maketitle

\section{Introduction}

In recent years, ideas and tools from the complex systems framework
have been extensively used to model various types of language dynamics,
such as language evolution (or how the structure of language evolves)
\cite{2011Steels}, language cognition (or the way in which the
human brain processes linguistic knowledge)~\cite{2007Edelman}, and
language competition (or the dynamics of language use in multilingual
communities)~\cite{2010Sole}. Particular emphasis has been given
in language competition, since a mass extinction of languages is anticipated
by linguists~\cite{1992Krauss}. The focus,
in understanding the factors that govern the competition of languages, lies in social
interactions. One of the main goals in the study of language competition
is to determine the mechanisms and conditions that lead either to
the coexistence of different languages or to the dominance
of a single language. This is of paramount importance for the design of
appropriate revitalization efforts and language planning policies.

Various approaches have been considered to model the dynamics of language
competition depending on the depth of societal description under consideration. For example, at a macroscopic level, where
only the population fractions speaking each language are considered,
there are approaches that make use of evolution equations borrowed from ecological
models~\cite{2005Mira} whereas others consider reaction-diffusion
equations~\cite{patriarca2009influence,2009Kandler}. These theoretical frameworks, amenable
to an analytical treatment, are able to describe the
observed trends in language growth or decline when compared to empirical
data. One of the pioneering works in this direction was put forward by
Abrams and Strogatz~\cite{2003Abrams} who studied the dynamics
of endangered languages. In this model, individuals in a binary-state society 
can speak either language A or B, allowing only for the possibility of societal bilingualism, i.e., the coexistence of two different monolingual groups~\cite{2006Appel}.
Subsequent generalizations of the original model considered the existence of bilingual groups~\cite{2005Wang}.

In general, the above approaches work well as long as individual
behavior can be neglected and one focuses on aggregated patterns.
If this is not the case, one should resort to a more microscopic description
of the system, considering approaches within the framework of agent-based
models (ABM)~\cite{2017Prochazka}. In this case, both the various individual characteristics that provide probabilities to switch to another language group and the detailed interactions between agents are taken into account, hence allowing for a description of the shifting mechanisms at an individual level. This permits a deeper understanding of how changes in different social factors can affect the dynamics of real-world systems~\cite{2018Williams}. Reference~\cite{2008Minett} focuses on ABMs in language competition models, exploring the effect of language status and education policies on the maintenance of a minority language. Another example would be Ref.~\cite{2013Castello}, which studies the dynamics of a community where two languages are used, each spoken by both monolingual and bilingual speakers. These works examine how language status and the individual likelihood to shift to another language impact language growth or decline. An alternative ABM framework is provided in Ref.~\cite{2012Patriarca}, which considers a game-theoretical approach to language competition modeling and discusses two different strategies encountered among minority-language speakers. 

Here, we build on an ABM suggested in Ref.~\cite{2016Carro}, which comprises a natural way of accounting for speakers who are potentially able to speak both languages. The main idea is that language is treated as a property of the interactions between individuals, rather than a property of the speaker. In this way, language serves as a means of communication, embodying distinct states for the links within the network. Unlike Ref.~\cite{2013Castello}, bilingualism is not an intermediate state but the result of individuals using different languages in different interactions. Furthermore, the node of the state can now encode the preference towards one of the languages, hence allowing for various degrees of bilingualism in the system. This preference in fact can be the result of a number of factors, such as the level of competence in that language or the degree of cultural attachment and affinity with the respective speech community. Our main contribution as compared to Ref.~\cite{2016Carro} is that in our model, as explained below, we allow for the possibility that speakers adapt the topology of their network of interactions if they are not satisfied with particular links (rewiring). Further, contrary to most rewiring mechanisms considered in the literature, we do not introduce a new global parameter for the propensity of the individuals to rewire, but rather this depends locally on the preference of each individual. 

Models that address the dynamics of link states have received increasing attention from different research areas such as social balance theory~\cite{2005Antal}, community detection~\cite{2009Traag}, network controllability~\cite{2012Nepusz}, and opinion formation~\cite{2019Saeedian}. Relevant to our work, Ref.~\cite{2012FernandezGracia} implements a majority rule for link states, while Ref.~\cite{2014Carro} develops a coevolution model that couples the aforementioned majority rule dynamics of link states with the evolution of the network topology. As previously discussed, in the context of language competition, Ref.~\cite{2016Carro} analyzes coupled node and link dynamics. Its main result is that, in contrast to most of the previously proposed models, the extinction of one of the languages is not an inevitable outcome of the dynamics. On the contrary, there is a wide range of possible asymptotic configurations, including not only extinction states but also frozen and dynamically trapped coexistence states, the probability of extinction being a decreasing function of the population size. Furthermore, recent studies consider the coupling of all three dynamics, i.e., nodes, links and network topology, as in Ref.~\cite{2020Saeedian}. Here, the link state is considered to be either attractive or repulsive, and it is found that repulsive interactions are a major cause of social polarization.

In this work, we assume that node and link dynamics are coupled in the same way as in Ref.~\cite{2016Carro} but we introduce a rewiring mechanism whose activation depends on the speakers' preferences. The main result of our work is that rewiring itself can contribute to the survival of both languages in the long time limit, with both languages having a significant presence. Here, we define survival as having over 5\% of the population speaking that language. We estimate the survival probability, the relative sizes of the two communities as well as that of the bilingual individuals (individuals that are using both languages), as a function of both the system size and the node update probability. We find that increasing the system size decreases the probability of language coexistence. The contrary holds true for the node update probability. However, by increasing either the system size or the node update probability, in those cases where the two languages end in a coexistence state in the long-time limit, the size of the two communities approaches each other, while the relative size of the community of bilinguals remains constant. This is a remarkable finding that is consistent with data taken from today's bilingual societies.
We also define a quantity, termed satisfaction, that captures the alignment of agents' preferences with their spoken language, and we see that, independently of the rewiring mechanism, agents appear almost always to be completely satisfied, except the bilinguals.

The work is organized as follows. In Sec.~\ref{sec:themodel} we present the model, explaining the node and link dynamics and how these are coupled. We also introduce the specifics of the rewiring mechanism and how these interfere with the topology of the network. The structural constraints imposed by the definition of the model as well as the particularities of the networks used for the numerical simulations are also described in this section. In Sec.~\ref{sec:results} we present the results of our studies in the long time limit. Finally, Sec.~\ref{conclusions} summarizes the main results of the work.

\section{The model}\label{sec:themodel}

We consider a population of $N$ speakers who are linguistically interacting among themselves. We represent this on a network, where nodes correspond to the speakers and links correspond to their interactions. Both nodes and links are characterized by state variables. The node state is a continuous variable $x_{i}\in[0,1]$ that represents the preference of speaker $i$ towards language A, ($1-x_{i}$ being their preference towards language B). Therefore, $x_i=1$ indicates an absolute or extreme preference for language A whereas $x_i=0$ an absolute or extreme preference for language B. The link state represents the language used in the interaction between agents and is described with a discrete binary variable $S_{ij}\in\{0,1\}$ such that $S_{ij}=1$ denotes the usage of language A between agents $i$ and $j$, and $S_{ij}=0$ corresponds to the usage of language B. 

States of nodes and links evolve asynchronously (single node or link update at each step) according to stochastic rules that we now describe. First, with probability $q$ a random node is chosen for an update, and with the complementary probability $1-q$ a random link is chosen instead. It should be noted that in the node selection, we assume a {\slshape link-update dynamics}, as described in~\cite{2005Suchecki}. This means that in order to select a random node, a link is selected first, and then one of the two nodes at the ends of the link is randomly selected. A fundamental difference of this procedure with traditional node-update schemes \cite{2006Castello,2012Patriarca}  is that nodes with a high degree are chosen more often to be updated. 

The probability $q$ sets the relationship between the evolution time scale of the speakers’ preferences and the time scale at which the language used in conversations changes. In the original version of the model without rewiring~\cite{2014Carro}, it was shown that the parameter $q$ simply affects the speed with which the asymptotic state is reached, but not its main features. As we later show, when rewiring is introduced the parameter $q$ does play a significant role. The posterior evolution depends on whether we have chosen to update a node or a link. 

As node and link updates are both present, there appears an ambiguity as to how to measure time in the simulations. Here we simply assume that the time unit, one Monte Carlo step, corresponds to $N$ updating attempts, irrespective of nodes or links.

\begin{figure*}
\begin{centering}
\includegraphics[scale=0.55]{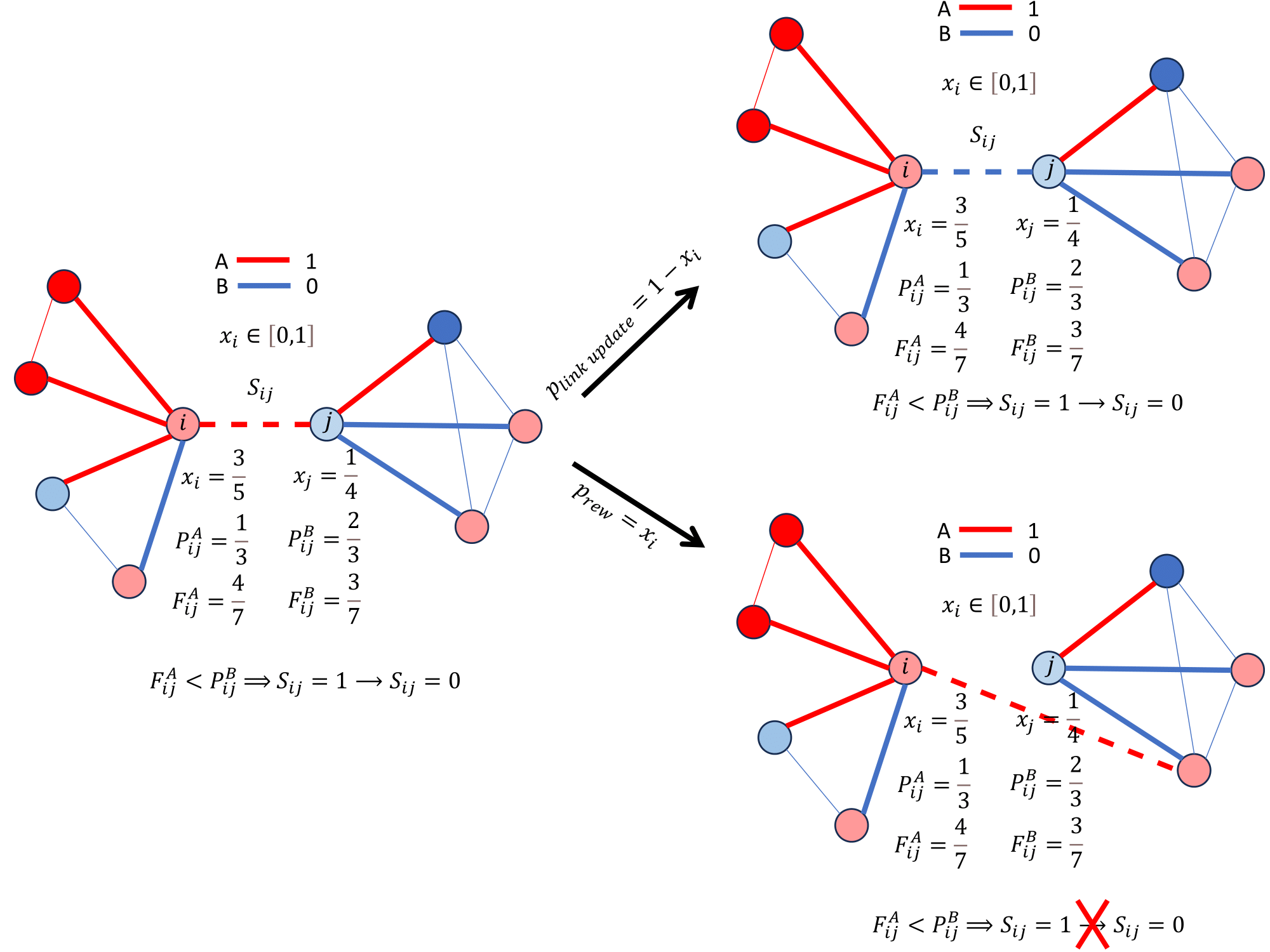}
\par\end{centering}
\caption{A schematic illustration of the rewiring mechanism. 
We present on the left-hand side a configuration where the selected link, in dashed line, should change its state according to the link state update algorithm described in \cite{2016Carro}. However, the novelty of our work is that we provide an alternative to this change. That is we allow for the most dissatisfied agent with the proposed change, agent $i$ in this example, to rewire to a neighbor of its current partner, with whom the link-update dynamics would not suggest a change of the state of the link. This rewiring happens with a probability equal to the preference of the most dissatisfied agent, which in the case of the scenario of this figure is equal to $x_i=\frac35$}
\label{fig:schematic}
\end{figure*}

\subsection{Network structure}

Recent studies have shown that real social networks are characterized
by a great number of triangles, resulting in high values of the clustering
coefficient~\cite{2003Newman,2003Dorogovtsev,2010Newman,2011Foster,2013ColomerDeSimon}. To fulfill this, we here consider networks with many triangles generated according to the algorithm proposed by Klimek and Thurner~\cite{2013Klimek}. This is a socially inspired algorithm and is based on the principle of triadic closure which states that individuals tend to make new acquaintances among friends of friends. There are three different mechanisms present in this algorithm: a) random link formation, b) triadic closure (link formation between nodes with a common neighbor) and c) node replacement (removal of a node with all its links and introduction of a new node with a certain number of links). With a suitable set of parameters, this model can reproduce data from a well-studied massive multiplayer online game~\cite{2010Szell,2010Szell-1,2012Szell,2016Klimek}. However, as in Ref.~\cite{2016Carro} we modify the aforementioned algorithm in order to guarantee that initially all nodes participate in at least one triangle. We also use the same parameter values found by Klimek and Thurner~\cite{2013Klimek} when calibrating their algorithm to the friendship network of the above-mentioned online game: a probability of triadic closure $c = 0.58$ ($1-c$ being the probability of random link formation) and a probability of node replacement $r = 0.12$. 

\begin{figure*}
\begin{centering}
\begin{tabular}{c}
\includegraphics[scale=0.675]{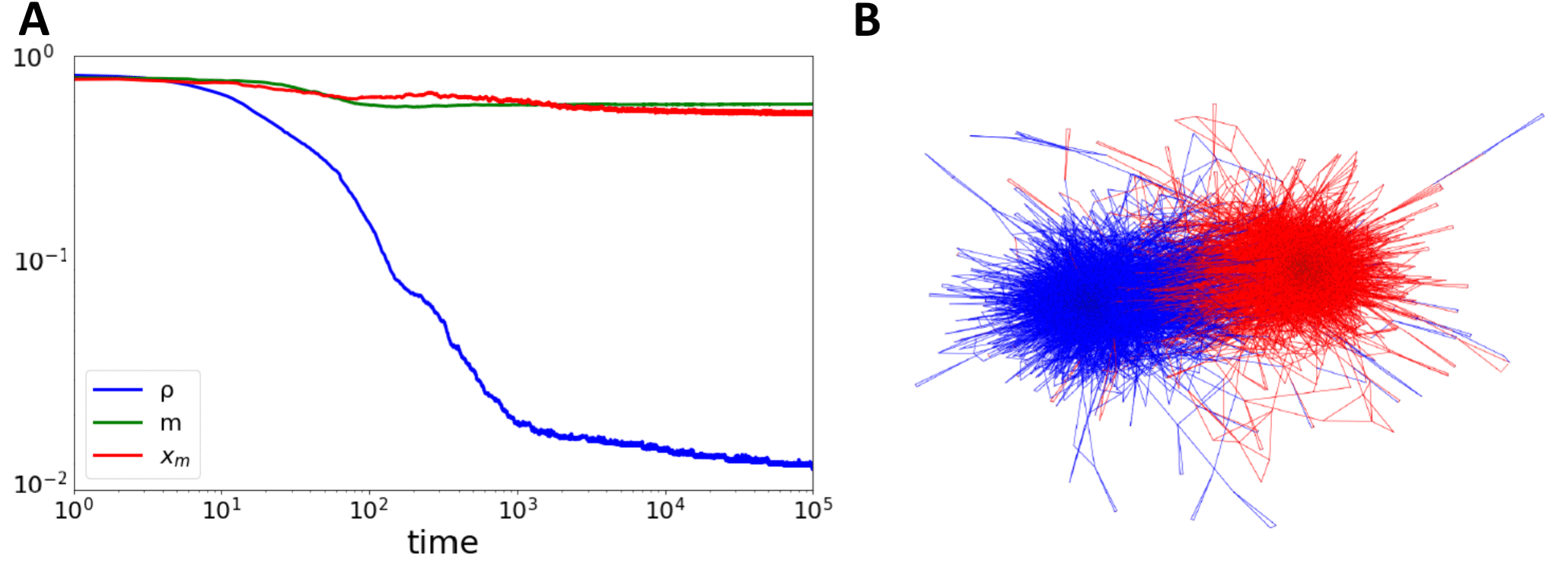}
\end{tabular}
\par\end{centering}
\caption{\label{fig:Asymptotic} \textbf{A}: Effect of rewiring on the order parameter $\rho$, the fraction of links in the minority language,
$m$, the average preference of the speakers for the minority
language, $x_m$, as well as on the final structure of the network. We see that the values
of $m$ and $x_{m}$ remain high (at the scale presented here the two are approximately equal and they overlap), while the value of $\rho$ decreases, indicating an increase of the number of nodes connected with links of the same language. At the same time, the fraction
of links in the minority language is not reduced, indicating that
the minority language comprises a significant fraction of the network.
Furthermore, the average preference of the speakers for the minority
language is not reduced. \textbf{B}: Network with rewiring in the long time limit. We
depict with red color the edges in the state $S_{ij}=1$
(language A) and with blue color the edges in the state $S_{ij}=0$ (language B). We observe a clear segregation between the
two communities, facilitating the coexistence of both languages. Parameters: $N=1000$, $q=0.5$, $t_\text{final}=10^5$. Results have been averaged over $100$ runs.}
\end{figure*}

\subsection{Evolution of node states}

The update of one node state, i.e., the preference of an agent for either language A or B is determined by the languages spoken by its neighboring agents during their interactions. After randomly selecting two of its neighbors engaged in communication, the agent adjusts its preference toward the language being utilized by this specific pair of neighbors. Effectively, we are assuming that group relationships take place on triangular social structures (see~\cite{2011Serrour} for a study on the relationship between communities and triangles). More precisely, when a node $i$ is chosen for updating, its state $x_{i}$ evolves according to the following rules:
\begin{enumerate}
 \item With probability $\frac{T_{i}^{A}}{T_{i}}$, set $x_{i}\rightarrow \min(x_{i}+\Delta x_i,1)$.
 \item With probability $1-\frac{T_{i}^{A}}{T_{i}}$, set $x_{i}\rightarrow \max(x_{i}-\Delta x_i,0)$.
\end{enumerate}
Here, $\Delta x_i=1/k_{i}$, with $k_{i}$ the degree of node $i$ (its total number of neighbors); $T_{i}$ is the total number of links between the neighbors of node $i$ and $T_{i}^{A}$ is the number of those links in state $1$, i.e., those in which language A is used. We should note that $\Delta x$ equals $1/k_{i}$ because we use a link-update scheme and nodes with many links are updated more often. Thus, these nodes also increase or decrease their state in smaller steps. Another justification for the form of $\Delta x$ is that nodes with fewer links tend to belong to fewer triangles and, therefore, each of them has a stronger influence on the node. 

\subsection{Evolution of link states}
Two factors contribute to the update of the language used between two
agents. First of all, we assume that the interaction between two given speakers tends to take place in the language most often used by both of them in their communications with other speakers. We postulate this as a way to encode the tendency of speakers to use as few languages as possible since the usage of more languages requires more cognitive effort~\cite{2001Jackson,2007Abutalebi}. For each link $i$--$j$ we define the majority pressure for language A, $F_{ij}^{A}$, as the fraction of
the number of interactions in which language A is used,
\begin{equation}
F_{ij}^{A}=\frac{k_{i}^{A}+k_{j}^{A}-2S_{ij}}{k_{i}+k_{j}-2},
\end{equation}
where $k_{i}^{A}$ stands for the number of neighbors in which
speaker $i$ uses language A, and $k_i$ is the total number of neighbors, the degree, of node $i$. A similar definition for language B, using $k_i^B$ as the number of neighbors in which
speaker $i$ uses language B, leads to $F_{ij}^B=1-F_{ij}^A$.

The second important factor that should be taken into account when
deciding the language used is the preference of speakers towards one language or the other. Combining the preferences
of both participants in each interaction $i$--$j$, we define the
link preference for language A as
\begin{equation}\label{eq_Pij}
P_{ij}^{A}=\begin{cases}
\dfrac{x_{i}x_{j}}{D}, & \text{if }\;D\ne0,\\
\dfrac{1}{2}, & \text{otherwise}.
\end{cases}
\end{equation}
with $D=x_{i}x_{j}+(1-x_{i})(1-x_{j})$.
Following Eq.~\eqref{eq_Pij}, when a link contains extreme preferences in both languages (either $x_{i}=0$ and $x_{j}=1$ or $x_{i}=1$ and $x_{j}=0$), the link preference is set to $1/2$. In addition, $P_{ij}^{A}$ yields $1$ if both agents prefer language A ($x_i=x_j=1$) and $0$ if both agents prefer language B ($x_i=x_j=0$). Furthermore, if one of the nodes is neutral with respect to language A, $x_{i}=1/2$, then the link preference is set to match the preference of the other node, $x_{j}$. Finally, the link preferences satisfy $P_{ij}^{A}\left(x_{i},x_{j}\right)=1-P_{ij}^{A}\left(1-x_{i},1 -x_{j}\right)$, reflecting the symmetry of the definition with respect to both languages. The definition for language B obeys $P_{ij}^{B}=1-P_{ij}^{A}$.

When a link $i$--$j$ is selected for updating, its new state is chosen according to the following rules:
\begin{enumerate}
 \item If the majority pressure for language A is larger than the link preference for language B, then language A is chosen.
 \item If the majority pressure for language A is smaller than the link preference for language B, then language B is chosen.
 \item If there is a tie between both languages, then one of them is randomly chosen.
\end{enumerate}
Leveraging the previously discussed symmetry between both languages, these rules can be mathematically expressed as
\begin{equation}
S_{ij}=\begin{cases}
1\;(\text{language A}), & \text{if } F_{ij}^{A}>P_{ij}^{B},\\ &\\
0\;(\text{language B}), & \text{if }F_{ij}^{A}<P_{ij}^{B},\\ &\\
0\;\text{or }1\;\text{randomly}, & \text{if }F_{ij}^{A}=P_{ij}^{B}.
\end{cases}\label{eq:transition_cond}
\end{equation}
Note that if all the speakers' preferences are fixed as $x_i=1/2,\,\forall i$, then all link preferences are also set to 1/2 and we recover the majority rule for link states analyzed in Ref.~\cite{2012FernandezGracia}: the state of a link is updated to the state of the majority of its neighboring links. With freely evolving preferences of the speakers, on the contrary, the threshold for a state to be considered a majority is not universal anymore and fixed at 1/2, but becomes local and dynamic. It is also noteworthy that, in a static network topology, speakers with extreme preferences ($x_{i}=0$ or $x_{i}=1$) impose their preferred language in all their conversations, except when they have to interact with a speaker with an extreme but opposite preference, in which case the language for their interaction is randomly selected. Thus, we are implicitly assuming that all speakers are able to use both languages, which makes our model suitable for bilingual societies. When a dynamic topology is considered, the speakers with extreme preferences do not necessarily impose their preferred language, since dissatisfied agents can instead rewire and form a new pair to use their preferred language.

\subsection{Rewiring dynamics}
We thus far have described the model developed in Ref.~\cite{2016Carro}. The main novelty of our work is the introduction of the possibility that
agents can change the local structure of their interactions in case of dissatisfaction.
For illustrative purposes, the dynamics which we now describe are depicted in Fig.~\ref{fig:schematic}.

When, as a result of the application of the conditions in Eq.~(\ref{eq:transition_cond}) the link is required to adapt its state, we allow for a link rewiring, maintaining the link's current state. The agent who retains the link is chosen based on being the one primarily dissatisfied with the expected change that results from the application of the conditions specified in Eq.~(\ref{eq:transition_cond}). For instance, if the proposed change was to switch the link state from 1 (language A) to 0 (language B), then the agent with the largest value of preference $x_{i}$ (i.e., with the largest preference towards language A) would be selected to rewire to another agent, say $\ell$, keeping the link state $S_{i\ell}=1$. 

This rewiring mechanism is probabilistic. In other words, either the proposed change of the state of the link is performed, or rewiring takes place. The latter occurs with a probability proportional to the preference of the most dissatisfied agent, i.e., the one who retains the link. The rewiring probability depends on the values of the nodes' preferences according to the following rules:
\begin{enumerate}
 \item If the proposed link update was $S_{ij}:0\rightarrow 1$, then set rewiring probability $\to1-\min\left(x_{i},x_{j}\right)$.
 \item If the proposed link update was $S_{ij}:1\rightarrow 0$, then set rewiring probability $\to\max\left(x_{i},x_{j}\right)$.
\end{enumerate}

The newly selected agent $\ell$ is randomly chosen from the set of neighbors of the neighbors of the agent exhibiting the highest level of dissatisfaction, say $i$, which are not themselves already neighbors of $i$. We also require that the new link $i$--$\ell$ satisfies the conditions given by Eq.~\eqref{eq:transition_cond}, i.e., $P^B_{i\ell}\le F^A_{i\ell}$ if $S_{i\ell}=1$, or $P^B_{i\ell}\le F^A_{i\ell}$ if $S_{i\ell}=0$. By choosing $\ell$ from the set of neighbors of neighbors of $i$ we ensure that, after rewiring, the number of triangles remains unaltered. However, if the proposed rewiring implies that the agent $j$ losing the link does not belong anymore to any triangle, then the rewiring is discarded. In this way, we adhere to the original constraint that all nodes must belong to at least one triangle. Finally, we note that, unlike the original model \cite{2016Carro}, the possible preference values are not defined as multiples of the inverse degree $1/k_{i}$. Hence, when rewiring is considered, the node preference remains unchanged even though its degree $k_i$ has changed. 

\begin{figure*}
\begin{centering}
\begin{tabular}{c}
\includegraphics[scale=0.8]{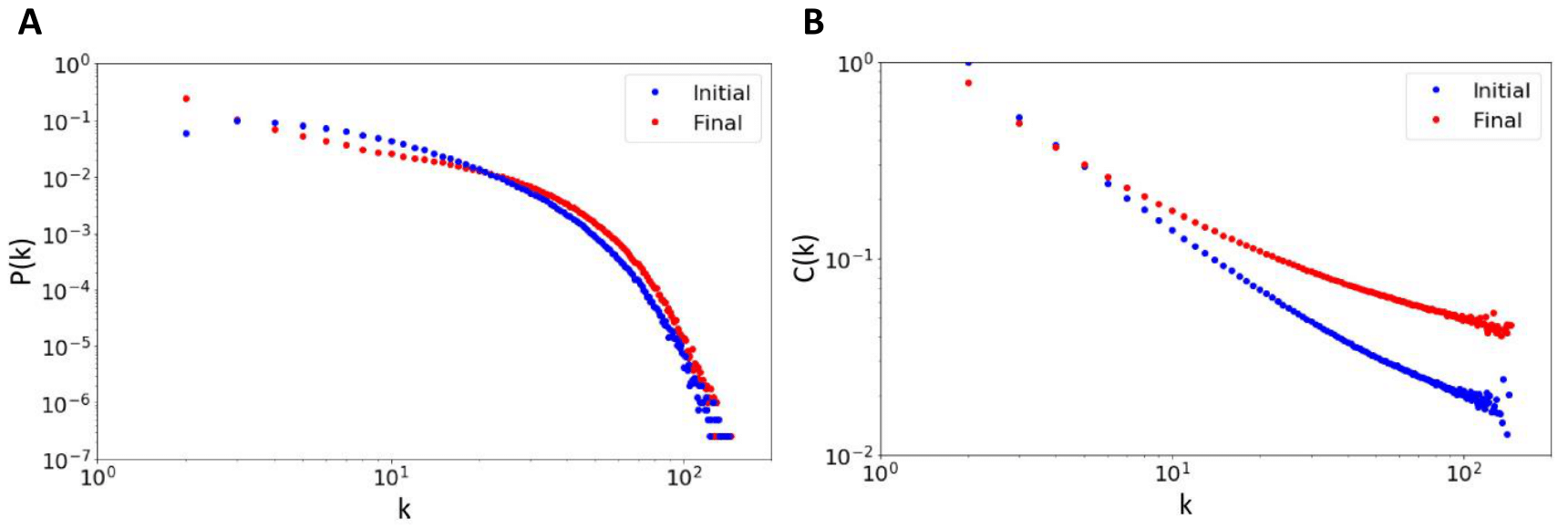} 
\end{tabular}
\par\end{centering}
\caption{Effect of rewiring on the degree distribution and average clustering coefficient. \textbf{A}: Initial and final degree distribution. No significant change is observed apart from a slight increase in the fraction of 2-degree nodes. \textbf{B}: Initial
and final average clustering coefficient as a function of the degree. We find that nodes with high degrees increase their clustering coefficient after the rewiring dynamics take place. Parameter values: $N=8000$, $q=0.5$, $t_\text{final}=10^5$. The data points are determined from an average of $100$ repetitions. }
\label{fig:rewiring_effects}
\end{figure*}

\section{Results}\label{sec:results}

Now we explore the convoluted dynamics of the system described above
through numerical simulations. In particular, we are interested in the asymptotic configurations reached after a long time. All our results reach a final time $t_\text{final}=10^5$ in units of Monte Carlo steps, although we have checked that larger times do not change significantly the main results for the system sizes analyzed here. When not stated explicitly, the network size considered is $N=1000$, the node update probability is set to $q=1/2$ and we average the results over a number of independent runs (typically $100$). 

Since a major goal is to examine the possibility of both languages surviving in the long term, we first consider the density of nodal interfaces~\cite{2012FernandezGracia,2014Carro}, defined as the fraction of pairs of connected links that show different states,
\begin{equation}
\rho=\frac{\sum_{i=1}^{N}k_{i}^{A}k_{i}^{B}}{\sum_{i=1}^{N}k_{i}\left(k_{i}-1\right)/2}.
\end{equation}
The order parameter $\rho$ is a measure of local order in the system, taking the value $\rho=0$ when all connected links share the same state and $\rho=1/2$ for a random distribution of link states. Alternatively, $\rho$ can also be seen as the usual density of active links (fraction of links connecting nodes with different states) when considering the line-graph of the original network~\cite{2012FernandezGracia,2014Carro,1965Rooij,1969Chartrand,2010MankaKrason,2011Krawczyk}.

We next introduce the fraction of links in the minority language
as a non-local measure characterizing the system
in terms of link states:
\begin{equation}\label{eq:m}
m=\dfrac{\min\left(\sum_{i}k_{i}^{A},\sum_{i}k_{i}^{B}\right)}{\sum_{i}k_{i}}.
\end{equation}
Minority language is obviously defined as the language that is used less in
the network. In this way, even if a population majority uses
a certain language in part of their interactions, we will still consider it to be the minority language if only a minority of the total number of communications actually takes place in that language. 

Finally, we characterize the system in terms of node states by
introducing the average preference of the speakers for the minority
language, 
\begin{equation}
x_{m}=\begin{cases}
N^{-1}\sum_{i}x_{i}, & \text{if }\sum_{i}k_{i}^{A}\leq\sum_{i}k_{i}^{B},\\ &\\
N^{-1}\sum_{i}\left(1-x_{i}\right), & \text{otherwise}.
\end{cases}
\end{equation}
This is a measure of the level of attachment for speakers of the minority language with their preferred language.

The time evolution of these three measures is presented in Fig.~\ref{fig:Asymptotic}A. All realizations start from a random initial distribution of states for both nodes and links. Regarding the links’ state, each link is assigned the value 0 or 1 with an equal probability, i.e. with probability 0.5. Regarding the agents’ preferences, each agent is assigned a preference value between 0 and 1 selected uniformly at random. As a consequence, the three measures $\rho$, $m$ and $x_m$ start from the value $0.5$. Then, the local order parameter $\rho$ monotonically decreases toward zero, which is a sign that most nodes are only connected to links of the same state. We checked the behavior of $\rho$ for a number of different values for the system size $N$ and the node update probability $q$ and it behaves approximately the same. The other two parameters, which are global in nature, for the specific simulation plotted here, remain constant. This is because the network splits into two approximately equal-sized communities, with each community speaking its own language. We confirm this by plotting the final state of the network in Fig.~\ref{fig:Asymptotic}\textbf{B}. This is in stark contrast with the results of Ref.~\cite{2016Carro}, where the authors find that in most cases a network of 1000 nodes either reaches the entire dominance of a single language or at most ghetto areas where one of the languages is spoken but as a minority language. We thus infer that the introduction of the rewiring mechanism significantly changes the final state of the system. In the next subsections, we examine this effect in more detail.

\subsection{Rewiring effect on network structure}

The main change in the network structure due to rewiring is shown in Fig.~\ref{fig:rewiring_effects}. We observe that the number of nodes with degree 2 increases with respect to its initial value, but the rest of the degree distribution remains basically unaltered by the rewiring. We also find that rewiring flattens the average clustering coefficient distribution. It follows from these observations that the rewiring effect increases the number of triangles in the network.

\begin{figure*}
\begin{centering}
\begin{tabular}{c}
\includegraphics[scale=0.55]{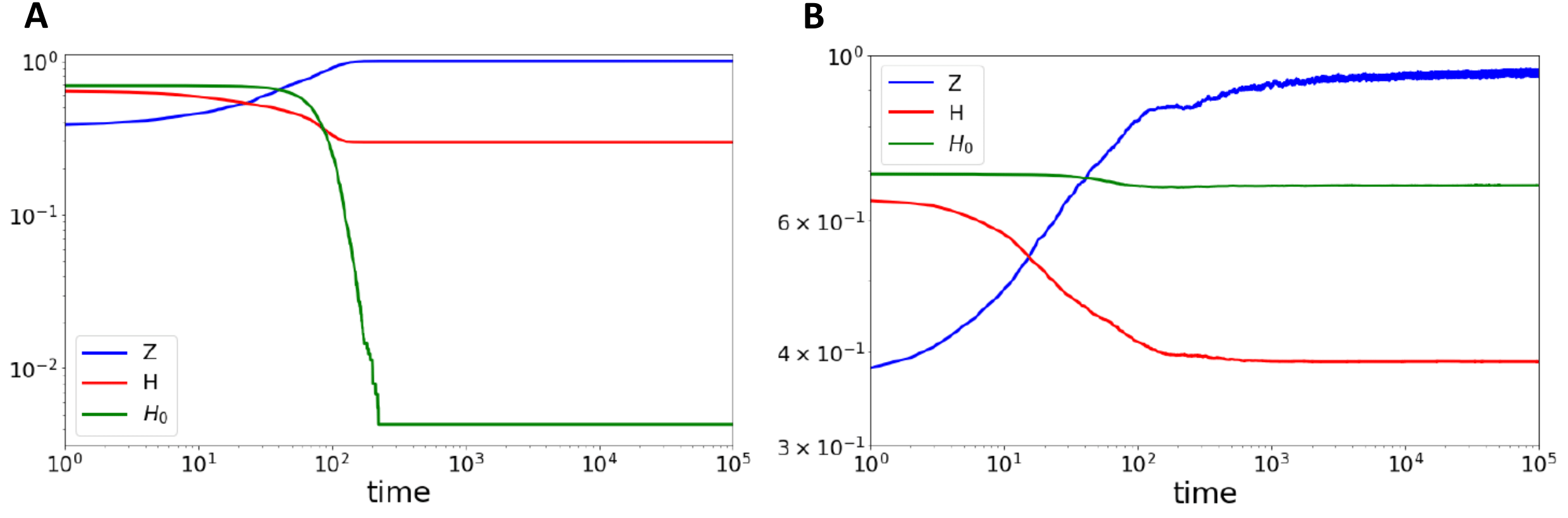} 
\end{tabular}
\par\end{centering}
\caption{\label{fig:Left:New-quantities} Effect of rewiring on the system entropy $H$, the null entropy $H_0$ and the satisfaction $Z$. \textbf{A}: No rewiring is allowed. We see that $Z$ grows up to 1 in the long-time limit. In addition, the value of $H$ decreases with time, until $H$ eventually becomes larger than the value of $H_0$, a sign of the existence of spatial mixing as expected. \textbf{B}: Rewiring is allowed. $Z$ reaches a value of $1$ but not exactly, a result of the existence of bilinguals
at the boundary of the two segregated communities. Furthermore, $H$ decreases similarly to the case without rewiring but the null entropy $H_{0}$ remains at much higher
values, which is a sign of spatial segregation. Parameters: $N=1000$, $q=0.5$, $t_\text{final}=10^5$.}
\end{figure*}

\subsection{Entropy and satisfaction}

To better understand the formation of linguistically polarized communities
due to rewiring in convoluted node and link dynamics, we now introduce the entropy,
which quantifies the spatial mixing or segregation in multigroup societies~\cite{2002Reardon,2021Louf}.
Segregation can be extracted from the entropy of an agent's interactions with a given group on average, compared to the entropy of the interactions of the whole population. The latter constitutes the null model, in which interactions in different languages are assigned anywhere on the network randomly and independently of the rest of the interactions of each individual. 

Let $H_{i}$ be the entropy associated with the states of the links attached to the $i$-th individual:
\begin{equation}\label{eq_Hi}
H_{i}=-\sum_{s\in\left\{ A,B\right\} }p\left(s|i\right)\log p\left(s|i\right),
\end{equation}
where 
\[
p\left(s|i\right)=\frac{k_{i}^{s}}{k_{i}}.
\]

We average Eq.~\eqref{eq_Hi} over all individuals,
\begin{equation}\label{eq_H}
H=\sum_{i=1,...,N}p\left(i\right)H_{i},
\end{equation}
where
\begin{equation}
p\left(i\right):=\frac{\sum_{s\in\left\{ A,B\right\}} k_{i}^{s}}
{\sum_{i,s}k_{i}^{s}}=\frac{k_{i}}{2L},
\end{equation}
with $L$ the total number of links.
Equation~\eqref{eq_H} is to be compared with the null
model, which assumes a non-segregated society:
\begin{equation}
H_{0}=-\sum_{s\in\left\{ A,B\right\} }\frac{L_{s}}{L}\log\left(\frac{L_{s}}{L}\right),
\end{equation}
where $L_{s}$ is the number of links at state $s$. This quantity, tends to the value 0 when one of the languages dominates over the other, while it's equal to $-log(0.5)$ when there is an equal number of links in language A and B in the network.  

We plot both $H$ and $H_0$ in Fig.~\ref{fig:Left:New-quantities}\textbf{B} and compare with the case of no rewiring in Fig.~\ref{fig:Left:New-quantities}\textbf{A}. We see that at the final state of the dynamics without rewiring the system entropy is higher than the value predicted with the null model. In addition, the value of the null model tends to 0, which shows that one language dominates over the other in the network. On the other hand, in Fig.~\ref{fig:Left:New-quantities}\textbf{B}, the entropy with rewiring is lower than that of the null model. In addition the null entropy's value is close to $-log(0.5)$. These indicate heterogeneity in the network, as a result of the fact that the network indeed splits into two approximately equal groups. The fact that the null entropy's value is close to $-\log(0.5)$ shows that there is a significant presence of both types of links in the network.

The network can be further characterized by calculating the average satisfaction of the agents:
\begin{equation}
Z=\frac{1}{2L}\sum_{i,j}\left(x_{i}+S_{ij}-1\right)^{2},
\end{equation}
where $\sum_{j}\left(x_{i}+S_{ij}-1\right)^{2}$ is understood as
a measure of the local satisfaction at node $i$ since this quantifies
the agreement between the local environment of the agent and its language preference.

Figure~\ref{fig:Left:New-quantities} also shows $Z$. We find that while the final satisfaction reaches its maximum value $1$ when no rewiring is allowed (see Figure ~\ref{fig:Left:New-quantities}\textbf{A}), the introduction of rewiring slightly
reduces the maximum value as can be seen in Fig.~\ref{fig:Left:New-quantities}\textbf{B}. This is a consequence of the existence of bilingual agents in the boundaries of the two groups. As these agents lie at the intersection of the two groups, their preferences as well as their links' states can fluctuate. As a consequence, the local environment and their preferences do not exactly match.

\subsection{Dependence on the node update probability}

\begin{figure}[t]
\begin{centering}
\begin{tabular}{c}
\includegraphics[scale=0.3]{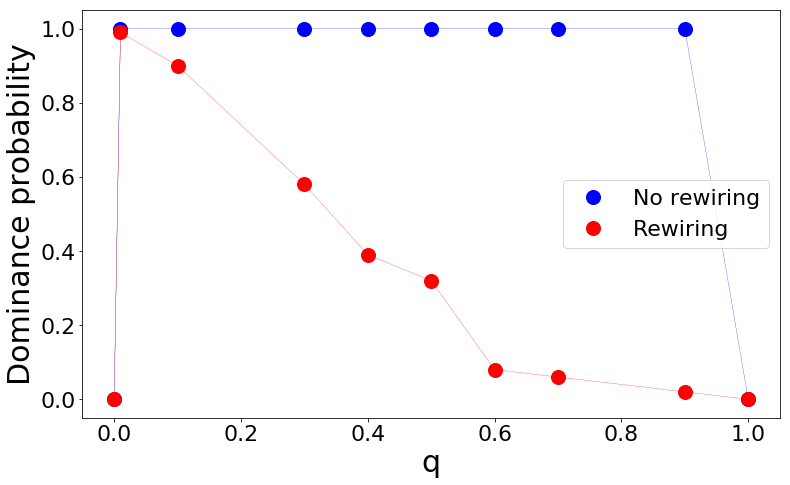} 
\end{tabular}
\par\end{centering}
\caption{\label{fig:ConVSq} Fraction of simulations that end up in the dominance state, i.e., where 95\% of the network agents only speak one language. Blue curve: no rewiring is allowed. The dominance scenario is fully present independently of $q\neq 0,1$. Red curve: rewiring is
allowed. Dominance probability decreases $q$ increases, vanishing when the update probability is high. 
As expected, in both cases the dominance probability vanishes for either $q=0$ or $q=1$. The parameters used were: N=1000, $t_\text{final}=10^5$ and 100 simulations are considered to calculate the probability.}
\end{figure}

 We recall that, despite the fact that all agents are able to speak both languages, an agent is considered bilingual only if it contains both A and B type links. The random initial conditions that we assumed above hence imply a high number of bilingual agents initially. In addition, as explained above, we define language dominance as the scenario where one of the languages, either A or B, is spoken by more than 95\% of the agents that use a single language. With this in mind, in Fig.~\ref{fig:ConVSq} we plot the probability that a simulation ends in a dominant state as a function of the node update parameter $q$. As we can observe, for the case without rewiring (blue dots), this probability remains unchanged as $q$ increases (except for the extreme cases $q=0$ and $q=1$. For the latter case, i.e. when the state of the links does not change, we see that we have a high number of bilingual agents, which is a consequence of the initial conditions we assume.). However, in the case where we do allow for rewiring (red dots), we obtain that this probability reduces as fewer updates of the link states occur. 

\begin{figure*}[]
\begin{centering}
\begin{tabular}{c}
\includegraphics[scale=0.55]{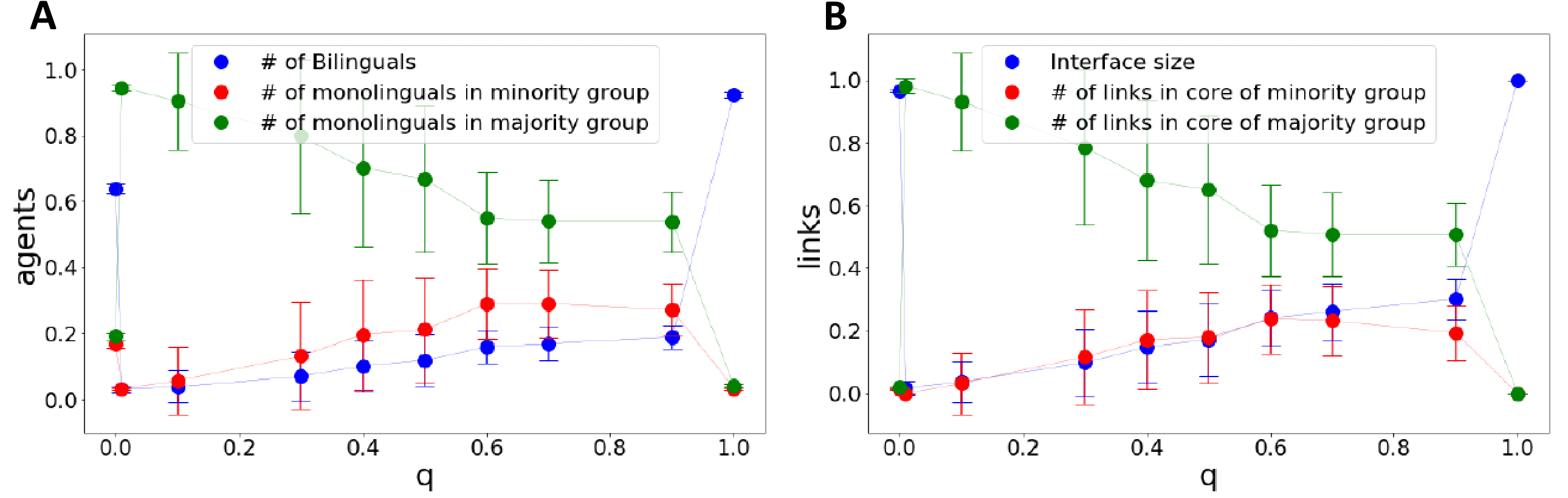}
\end{tabular}
\par\end{centering}
\caption{\label{fig:q-dependence} Language group populations as a function of the node update parameter $q$. For the extreme values of $q=0$ and $q=1$, i.e., when
only links are updated or when only nodes are updated, the network
is mixed. In other words, bilingual agents prevail. As soon as a mixture of
updates is allowed ($0<q<1$), we see the appearance
of a majority group. The network reaches a polarized state (segregation). Furthermore, we see that
the number of bilingual agents \textbf{(A)} as well as that of interface links \textbf{(B)} increases as $q$ increases. At the same time, the sizes of the
minority and majority groups approach each other. The displayed error bars correspond to one standard deviation. Parameters: N=1000, $t_\text{final}=10^5$.}
\end{figure*}

In Fig.~\ref{fig:q-dependence}, we study the final form of the system depending on the parameter $q$, for those networks in which language dominance is not reached.
We now focus on the case with rewiring only, since the case without rewiring showed no significant changes in Fig.~\ref{fig:ConVSq}.
We find that for the extreme values $q=0$ and $q=1$,
the network is mostly composed of bilingual agents, although for $q=1$
the number of bilingual agents is significantly higher
than for the $q=0$ case. For all other values of $q$, we find that in the absence of rewiring the network is characterized by the dominance
of one of the two languages, whereas when we allow for rewiring the network is polarized. Furthermore, increasing $q$ causes a decrease in the difference between the majority and minority group sizes. We find that the ensemble average preference for all groups of agents (the bilinguals, the majority language, and
the minority language groups) fluctuates around 0.5 as expected from the symmetry of our model with regard to both languages. We understand this dependence on $q$ as follows. When $q$ increases the link probability also increases but in turn this leads to more rewiring, which in the end favors coexistence. Then, we expect that in more dynamic societies (e.g., urban regions) the linguistic scenario will be more heterogeneous with the presence of both languages unlike static societies (e.g., rural regions), where one of the languages is more likely to disappear. This is consistent with the usage of linguistic varieties recently observed in social media~\cite{goncalves}. Finally, we should note that initial conditions with lower percentage of bilingual agents were also examined. For example we considered the case of starting with an already segregated society in terms of links' state (which implies a low number of bilinguals), but maintaining the uniformly random distribution of preferences and the results remained largely unchanged. We also considered other scenarios of intermediate number of bilinguals and we verified the robustness of our results.


\subsection{Dependence on system size}

\begin{figure}[t]
\begin{centering}
\begin{tabular}{c}
\includegraphics[scale=0.3]{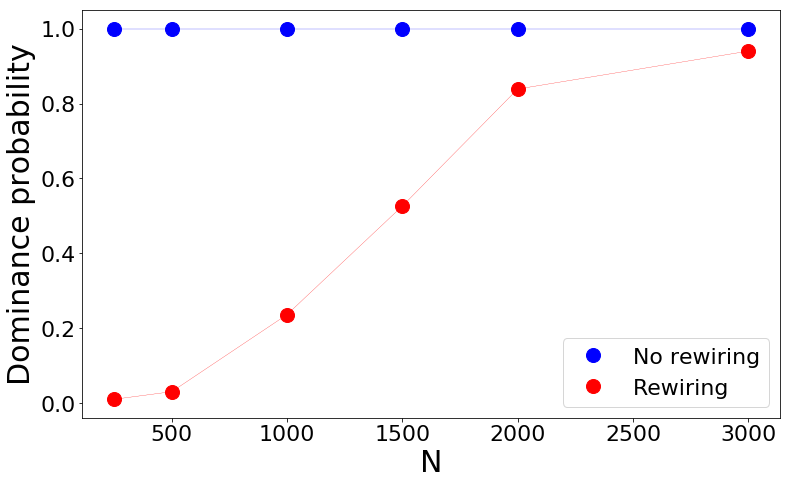} 
\end{tabular}
\par\end{centering}
\caption{\label{fig:consensusVSsystem-size} Effect of rewiring on the dependence of the dominance probability on system size. Blue curve: rewiring is not allowed. As the system size increases the probability of dominance is constant and always equal to 1. Red curve: rewiring is allowed. As the system size
increases it becomes more and more likely that one language dominates in the
long run. Parameters: $q=0.5$, $t_\text{final}=10^5$ and 100 realizations.}
\end{figure}

We now calculate the probability of the simulation reaching dominance
as a function of the system size. Our results are depicted in Fig.~\ref{fig:consensusVSsystem-size}.
We find that as the system size increases it is highly likely that the final state of the network reaches language dominance (red dots). To compare with, we also show
the results when no rewiring is allowed (blue dots). In this case, the dominance probability is independent of the system size. Therefore, rewiring is key in determining the fate
of the network. In particular, for small systems coexistence is the most likely
scenario. For large systems, however, rewiring is less effective for the maintenance
of a language.

\begin{figure*}[]
\begin{centering}
\begin{tabular}{c}
\includegraphics[scale=0.55]{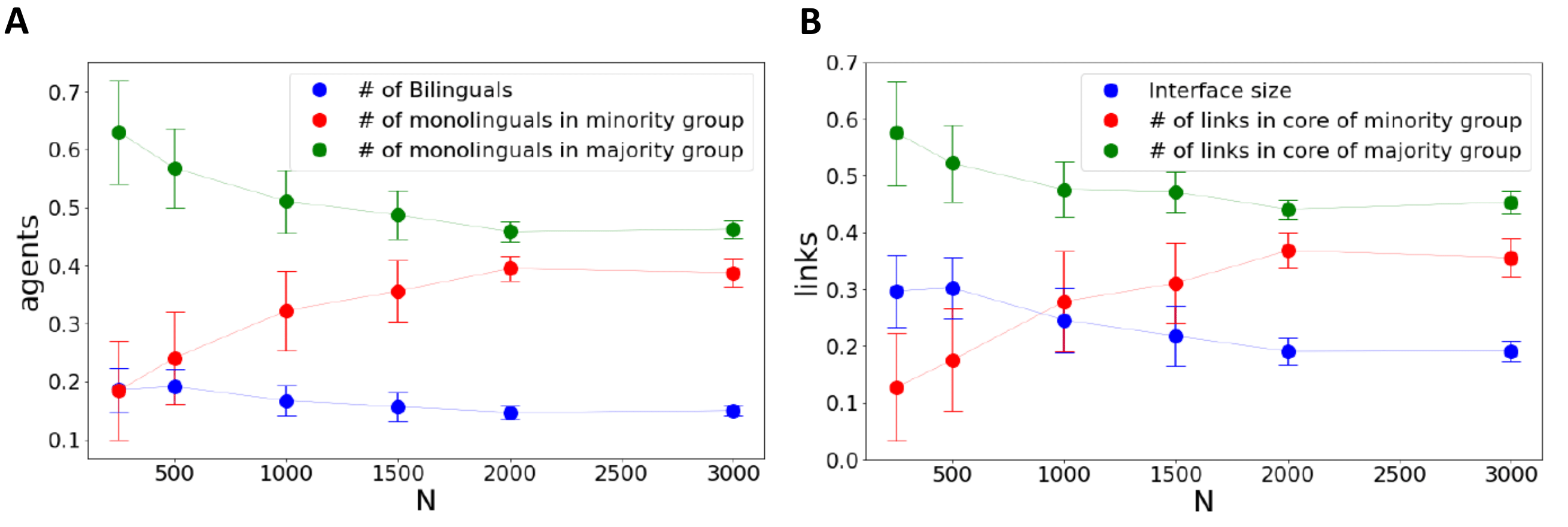} 
\end{tabular}
\par\end{centering}
\caption{\label{fig:system-size} Effect of system size on the proportions of bilingual, minority and majority speakers when rewiring is permitted. \textbf{A}: fraction of agents belonging
to each group as a function of the system size. \textbf{B}:
fraction of links belonging to each group as a function
of the system size. As we can see in both figures, as the system size
increases the minority and majority groups become
almost equal in size. The displayed error bars correspond to one standard deviation. Parameters: $q=0.5$, $t_\text{final}=10^5$.}
\end{figure*}

In Fig.~\ref{fig:system-size}, we study the dependence of the final
state on the system size $N$. Our simulations show that as the system size increases
the proportions of the majority and minority communities approach each other, while 
the magnitude of the interface constitutes roughly the same fraction of the entire network, independently of the system size. This is reflected in both individuals
(see Fig.~\ref{fig:system-size}\textbf{A}) and their connections (see Fig.~\ref{fig:system-size}\textbf{B}). We remark that the most likely situation for $N$ large is the extinction of one of the languages. Yet, in the scenario that both languages manage to coexist in a large region, the communities speaking these languages would approximately have the same sizes. Furthermore, the magnitude of the bilingual group of the region would remain approximately constant. This is a remarkable finding that deserves further attention.

\begin{figure}[t]
\begin{centering}
\begin{tabular}{c}
\includegraphics[scale=0.21]{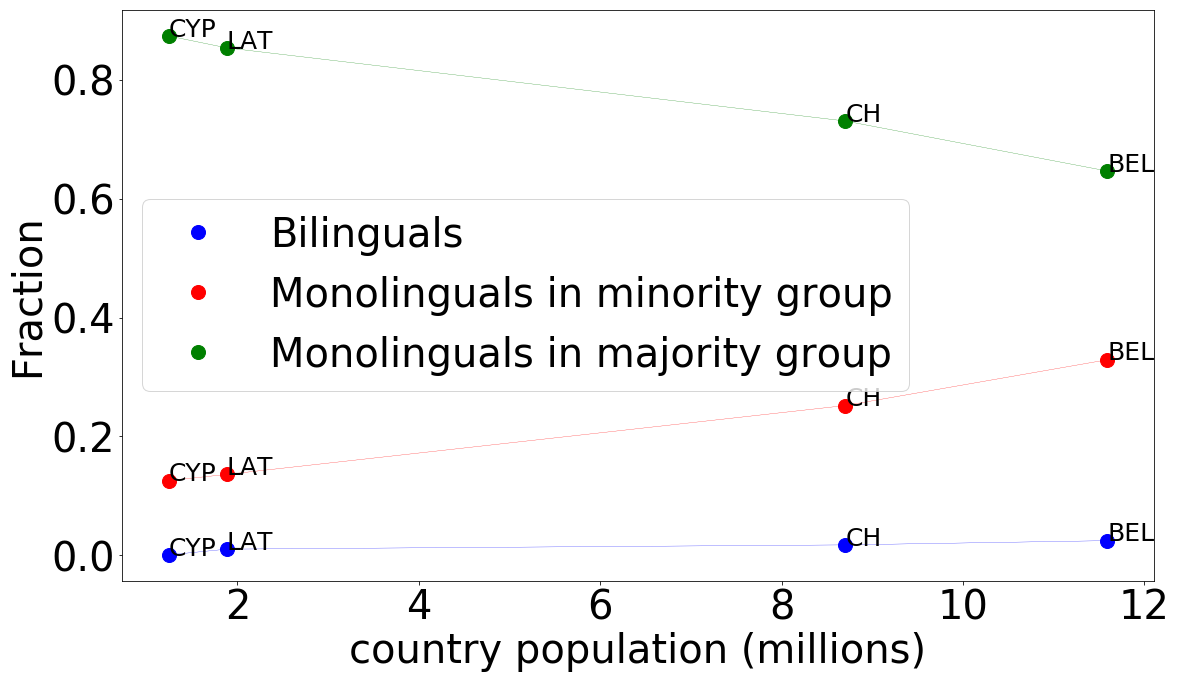}\tabularnewline
\end{tabular}
\par\end{centering}
\caption{\label{fig:system-size-COUNTRIES} Proportion of majority and minority monolinguals as well as bilinguals for a number of bilingual (mostly segregated) countries: Cyprus (CYP), Latvia (LAT), Switzerland (CH), and Belgium (BEL). Data taken from Ref.~\cite{2021Louf}.}
\end{figure}

To examine whether this is indeed observed in real life, we consider a number of bilingual countries and regions showing a clear segregation and plot the fraction of speakers of both the majority and minority languages as well as the percentage of bilinguals. We obtain these data from geolocalized posts in Twitter with automatic language detection~\cite{2021Louf}. We find that for a number of countries studied there is a qualitative agreement between the data and our findings, as can be seen in Fig.~\ref{fig:system-size-COUNTRIES}. In particular, we observe that from the populations of the bilingual countries of Cyprus, Latvia, Switzerland (ignoring the Italian-speaking population) and Belgium, the sizes of the majority and minority groups approach each other for bigger countries. Moreover, we find that the percentage of bilingual individuals remains constant, independently of the country size, also in agreement with our numerical predictions discussed above. 


\section{Conclusions}\label{conclusions}

We have studied an agent-based model for language dynamics valid for bilingual societies,  enabling rewiring in the link dynamics. This models the agents' decision to continue speaking their preferred language while engaging with a different agent. As a consequence of this, we find that both languages can coexist in the long-term state of the network, resulting in a bipolarized network. As the system size increases, the probability that a single language dominates also increases. Surprisingly enough, for those simulations that show language coexistence, we find that the relative size of the bilingual agents remains constant with the system size, while the difference between the majority and the minority language groups decreases. We compare this with population data and language usage in bilingual countries and obtain good agreement. Another key quantity is the likelihood of individuals updating their language preference based on the languages spoken by their neighbors. Increasing this likelihood also strengthens the probability of reaching language coexistence, thereby reducing language shift. 

Clearly, our work implies that understanding the dynamics of social network interactions has a profound significance for language planning~\cite{Kaplan1997} and the design of adequate policies that promote the revitalization of endangered languages~\cite{sallabank2013attitudes}. Our model assumes that language is a property of the interaction rather than a property of the speaker and that bilingualism (or multilingualism) is a result of language use. Of course, individuals have a language preference and this is modeled with a continuous parameter. But, importantly, agents can also adapt the topology of their interactions to align with their preferences.

Further extensions of the model should take into account not only linguistically segregated regions but also mixed societies, especially when the interlinguistic similarity is high~\cite{mira2011importance}. Another important issue is demographics. In this case, networks of varying sizes should be treated dynamically. It would be interesting to apply the formalism of multilayer networks~\cite{de2023more} to account for heterogeneous speech communities and how layer interdependence affects language coexistence, possibly incorporating social factors such as economic class~\cite{louf2023dialects} and status~\cite{rosillo2023modelling}. Furthermore, one could also consider the role of memory in the dynamics of both the nodes and the links as in Ref.~\cite{2023Scialla}.
In addition, one could consider the case where the dynamics, e.g. of the language usage decision, are defined taking into account group interactions rather than simply pairwise interactions \cite{2013Perc}. 
Finally, we could also consider probabilistic updates for the link choice as well, i.e., the introduction of noise. Currently, for language shift to take place it suffices that the majority pressure for the other language is larger than the link preference for the currently used language, but we could assume a probabilistic change that would depend on the difference between the two quantities. This would describe the scenario of a not completely rational choice of the agents when taking into account the two sociological mechanisms, the majority pressure and the link preference, considered in this work. 

\acknowledgments{Partial financial support has been received from the Agencia Estatal de Investigaci\'on (AEI, MCI, Spain) MCIN/AEI/
10.13039/501100011033 and Fondo Europeo de Desarrollo Regional (FEDER, UE) under Project APASOS (PID2021-122256NB-C21), the María de Maeztu Program for units of Excellence in R\&D, under grant CEX2021-001164-M. and by the Government of the Balearic Islands CAIB fund ITS2017-006 under project CAFECONMIEL (PDR2020/51). C.C. was supported by Direcció General de Política Universitària i Recerca from the government of the Balearic Islands through the postdoctoral program Margalida Comas.}

\bibliographystyle{unsrt}
\bibliography{Language}

\begin{thebibliography}{10}

\bibitem{2011Steels}
L.~Steels.
\newblock Modeling the cultural evolution of language.
\newblock {\em Physics of Life Reviews}, 8(4):339--356, 2011.

\bibitem{2007Edelman}
S.~Edelman and H.~Waterfall.
\newblock Behavioral and computational aspects of language and its acquisition.
\newblock {\em Physics of Life Reviews}, 4(4):253--277, 2007.

\bibitem{2010Sole}
R.~V. Sole, B.~Corominas-Murtra, and J.~Fortuny.
\newblock Diversity, competition, extinction: the ecophysics of language
  change.
\newblock {\em J. R. Soc. Interface.}, 7:1647–1664, 2010.

\bibitem{1992Krauss}
M.~Krauss.
\newblock The world's languages in crisis.
\newblock {\em Language}, 68, 1992.

\bibitem{2005Mira}
J.~Mira and Á. Paredes.
\newblock Interlinguistic similarity and language death dynamics.
\newblock {\em Europhysics Letters}, 69(6):1031, 2005.

\bibitem{patriarca2009influence}
M.~Patriarca and E.~Heinsalu.
\newblock Influence of geography on language competition.
\newblock {\em Physica A: Statistical Mechanics and its Applications},
  388(2-3):174--186, 2009.

\bibitem{2009Kandler}
A.~Kandler.
\newblock Demography and language competition.
\newblock {\em Human Biology}, 81(2/3):181--210, 2009.

\bibitem{2003Abrams}
D.~M. Abrams and S.~H. Strogatz.
\newblock Modelling the dynamics of language death.
\newblock {\em Nature}, 424(6951), 2003.

\bibitem{2006Appel}
R.~Appel and P.~Muysken.
\newblock {\em Language Contact and Bilingualism}.
\newblock Cambridge University Press, 2006.

\bibitem{2005Wang}
W.~S.~Y. Wang and J.~W. Minett.
\newblock The invasion of language: emergence, change and death.
\newblock {\em Trends Ecol. Evol.}, 20(263):263--269, 2005.

\bibitem{2017Prochazka}
K.~Prochazka and G.~Vogl.
\newblock Quantifying the driving factors for language shift in a bilingual
  region.
\newblock {\em Proceedings of the National Academy of Sciences},
  114(17):4365--4369, 2017.

\bibitem{2018Williams}
R.~A. Williams.
\newblock Lessons learned on development and application of agent-based models
  of complex dynamical systems.
\newblock {\em Simulation Modelling Practice and Theory}, 83:201--212, 2018.
\newblock Agent-based Modelling and Simulation.

\bibitem{2008Minett}
J.~W. Minett and W.~S-Y. Wang.
\newblock Modelling endangered languages: The effects of bilingualism and
  social structure.
\newblock {\em Lingua}, 118(1):19--45, 2008.

\bibitem{2013Castello}
X.~Castello, L.~Loureiro-Porto, and M.~San Miguel.
\newblock Agent-based models of language competition.
\newblock {\em International Journal of the Sociology of Language},
  2013(221):21--51, 2013.

\bibitem{2012Patriarca}
M.~Patriarca, X.~Castello, J.~R. Uriarte, V.~M. Eguiluz, and M.~San Miguel.
\newblock Modelling two-language competition dynamics.
\newblock {\em Advances in Complex Systems}, 15(03n04):1250048, 2012.

\bibitem{2016Carro}
A.~Carro, R.~Toral, and M.~San Miguel.
\newblock Coupled dynamics of node and link states in complex networks: a model
  for language competition.
\newblock {\em New Journal of Physics}, 18(11):113056, 2016.

\bibitem{2005Antal}
T.~Antal, P.~L. Krapivsky, and S.~Redner.
\newblock Dynamics of social balance on networks.
\newblock {\em Phys. Rev. E}, 72:036121, 2005.

\bibitem{2009Traag}
V.~A. Traag and J.~Bruggeman.
\newblock Community detection in networks with positive and negative links.
\newblock {\em Phys. Rev. E}, 80:036115, 2009.

\bibitem{2012Nepusz}
T.~Nepusz and T.~Vicsek.
\newblock Controlling edge dynamics in complex networks.
\newblock {\em Nature Physics}, 8:568--573, 2012.

\bibitem{2019Saeedian}
M.~Saeedian, M.~San Miguel, and R.~Toral.
\newblock {Absorbing phase transition in the coupled dynamics of node and link
  states in random networks}.
\newblock {\em Scientific Reports}, 9(1):9726, 2019.

\bibitem{2012FernandezGracia}
J.~Fern\'andez-Gracia, X.~Castello, V.~M. Eguiluz, and M.~San Miguel.
\newblock Dynamics of link states in complex networks: The case of a majority
  rule.
\newblock {\em Phys. Rev. E}, 86:066113, 2012.

\bibitem{2014Carro}
A.~Carro, F.~Vazquez, R.~Toral, and M.~San Miguel.
\newblock Fragmentation transition in a coevolving network with link-state
  dynamics.
\newblock {\em Phys. Rev. E}, 89:062802, 2014.

\bibitem{2020Saeedian}
M.~Saeedian, M.~San Miguel, and R.~Toral.
\newblock Absorbing-state transition in a coevolution model with node and link
  states in an adaptive network: network fragmentation transition at
  criticality.
\newblock {\em New Journal of Physics}, 22(11):113001, 2020.

\bibitem{2005Suchecki}
K.~Suchecki, V.~M. Eguíluz, and M.~San Miguel.
\newblock Conservation laws for the voter model in complex networks.
\newblock {\em Europhysics Letters}, 69(2):228, 2004.

\bibitem{2006Castello}
X.~Castelló, V.~M. Eguíluz, and M.~San Miguel.
\newblock Ordering dynamics with two non-excluding options: bilingualism in
  language competition.
\newblock {\em New Journal of Physics}, 8(12):308, 2006.

\bibitem{2003Newman}
M.~E.~J. Newman and J.~Park.
\newblock Why social networks are different from other types of networks.
\newblock {\em Phys. Rev. E}, 68:036122, 2003.

\bibitem{2003Dorogovtsev}
S.~Dorogovtsev and J.~F.~F. Mendes.
\newblock {\em Evolution of Networks: From Biological Nets to the Internet and
  WWW}.
\newblock Oxford Academic, 2003.

\bibitem{2010Newman}
M.~E.~J. Newman.
\newblock {\em Networks: An Introduction}.
\newblock Oxford University Press, 2010.

\bibitem{2011Foster}
D.~V. Foster, J.~G. Foster, P.~Grassberger, and M.~Paczuski.
\newblock Clustering drives assortativity and community structure in ensembles
  of networks.
\newblock {\em Phys. Rev. E}, 84:066117, 2011.

\bibitem{2013ColomerDeSimon}
P.~Colomer de~Simón, M.~A. Serrano, M.~G. Beiró, J.~I. Alvarez-Hamelin, and
  M.~Boguñá.
\newblock Deciphering the global organization of clustering in real complex
  networks.
\newblock {\em Scientific Reports}, 3:2517, 2013.

\bibitem{2013Klimek}
P.~Klimek and S.~Thurner.
\newblock Triadic closure dynamics drives scaling laws in social multiplex
  networks.
\newblock {\em New Journal of Physics}, 15(6):063008, 2013.

\bibitem{2010Szell}
M.~Szell, R.~Lambiotte, and S.~Thurner.
\newblock Multirelational organization of large-scale social networks in an
  online world.
\newblock {\em Proceedings of the National Academy of Sciences},
  107(31):13636--13641, 2010.

\bibitem{2010Szell-1}
M.~Szell and S.~Thurner.
\newblock Measuring social dynamics in a massive multiplayer online game.
\newblock {\em Social Networks}, 32(4):313--329, 2010.

\bibitem{2012Szell}
M.~Szell and S.~Thurner.
\newblock Social dynamics in a large-scale online game.
\newblock {\em Advances in Complex Systems}, 15(06):1250064, 2012.

\bibitem{2016Klimek}
P.~Klimek, M.~Diakonova, V.~M. Eguiluz, M.~San Miguel, and S.~Thurner.
\newblock {Dynamical origins of the community structure of an online
  multi-layer society}.
\newblock {\em New Journal of Physics}, 18(8):83045, 2016.

\bibitem{2011Serrour}
B.~Serrour, A.~Arenas, and S.~Gomez.
\newblock Detecting communities of triangles in complex networks using spectral
  optimization.
\newblock {\em Computer Communications}, 34(5):629--634, 2011.
\newblock Special Issue: Complex Networks.

\bibitem{2001Jackson}
G.~M. Jackson, R.~Swainson, R.~Cunnington, and S.~R. Jackson.
\newblock Erp correlates of executive control during repeated language
  switching.
\newblock {\em Bilingualism: Language and Cognition}, 4(2):169–178, 2001.

\bibitem{2007Abutalebi}
J.~Abutalebi and D.~Green.
\newblock Bilingual language production: The neurocognition of language
  representation and control.
\newblock {\em Journal of Neurolinguistics}, 20(3):242--275, 2007.

\bibitem{1965Rooij}
A.~van Rooij and H.~S. Wilf.
\newblock The interchange graph of a finite graph.
\newblock {\em Acta Mathematica Academiae Scientiarum Hungarica}, 16:263--269,
  1965.

\bibitem{1969Chartrand}
G.~Chartrand and M.~J. Stewart.
\newblock The connectivity of line-graphs.
\newblock {\em Mathematische Annalen}, 182:170--174, 1969.

\bibitem{2010MankaKrason}
A.~Mańka-Krasoń, A.~Mwijage, and K.~Kułakowski.
\newblock Clustering in random line graphs.
\newblock {\em Computer Physics Communications}, 181(1):118--121, 2010.

\bibitem{2011Krawczyk}
M.~J. Krawczyk, L.~Muchnik, A.~Mańka-Krasoń, and K.~Kułakowski.
\newblock {Line graphs as social networks}.
\newblock {\em Physica A: Statistical Mechanics and its Applications},
  390(13):2611--2618, 2011.

\bibitem{2002Reardon}
S.~F. Reardon and G.~Firebaugh.
\newblock Measures of multigroup segregation.
\newblock {\em Sociological methodology}, 32:33--67, 2002.

\bibitem{2021Louf}
T.~Louf, D~S{\'a}nchez, and J.~Ramasco.
\newblock Capturing the diversity of multilingual societies.
\newblock {\em Phys. Rev. Res.}, 3:043146, 2021.

\bibitem{goncalves}
Bruno Gon{\c{c}}alves and David S{\'a}nchez.
\newblock Crowdsourcing dialect characterization through twitter.
\newblock {\em PloS one}, 9(11):e112074, 2014.

\bibitem{Kaplan1997}
R.~B. Kaplan and R.~B. Baldauf.
\newblock {\em Language planning from practice to theory}, volume 108.
\newblock Multilingual Matters, 1997.

\bibitem{sallabank2013attitudes}
J.~Sallabank.
\newblock {\em Attitudes to endangered languages: Identities and policies}.
\newblock Cambridge University Press, 2013.

\bibitem{mira2011importance}
J.~Mira, L.~F. Seoane, and J.~J. Nieto.
\newblock The importance of interlinguistic similarity and stable bilingualism
  when two languages compete.
\newblock {\em New Journal of Physics}, 13(3):033007, 2011.

\bibitem{de2023more}
M.~De Domenico.
\newblock More is different in real-world multilayer networks.
\newblock {\em Nature Physics}, pages 1--16, 2023.

\bibitem{louf2023dialects}
T.~Louf, J.~J. Ramasco, D.~S{\'a}nchez, and M.~Karsai.
\newblock When dialects collide: How socioeconomic mixing affects language use.
\newblock {\em arXiv preprint arXiv:2307.10016}, 2023.

\bibitem{rosillo2023modelling}
P.~Rosillo-Rodes, M.~San Miguel, and D.~S{\'a}nchez.
\newblock Modelling language ideologies for the dynamics of languages in
  contact.
\newblock {\em arXiv preprint arXiv:2307.02845}, 2023.

\bibitem{2023Scialla}
S.~Scialla, J.-K. Liivand, M.~Patriarca, and E.~Heinsalu.
\newblock A three-state language competition model including language learning
  and attrition.
\newblock {\em Frontiers in Complex Systems}, 1:1266733, 2023.

\bibitem{2013Perc}
M.~Perc, J.~Gómez-Gardeñes, A.~Szolnoki, L.~M. Floría, and Y.~Moreno.
\newblock Evolutionary dynamics of group interactions on structured
  populations: a review.
\newblock {\em J. R. Soc. Interface}, 10(80):20120997, 2013.

\end{thebibliography}

\end{document}